%% file: main.tex
\RequirePackage{lineno}
\documentclass[preprint,superscriptaddress,aps,prl,superscriptaddress,showpacs,preprintnumbers,amsmath,amssymb]{revtex4}
\usepackage{color}
\usepackage{longtable}
\usepackage{graphicx}
\usepackage{dcolumn}  
\usepackage{pdflscape}
\usepackage{amsmath}
\usepackage{lineno}
\usepackage{xspace}
\usepackage{siunitx}
\usepackage{orcidlink}

\def\PRL{ Phys. Rev. Lett.}
\def\PRD{{ Phys. Rev.} D}

\def\Mbc{\ensuremath{M^{}_{\rm bc}}}
\def\mev{\ensuremath{\mathrm{\,Me\kern -0.1em V}}\;}
\def\Y#1S{\ensuremath{\Upsilon{(#1S)}}\xspace}
\def\FourS {\Y4S}
\newcommand{\gevc}{\ensuremath{{\mathrm{\,Ge\kern -0.1em V\!/}c}}\xspace}
\def\Bu{\ensuremath{B^+}\xspace}
\def\Bp{\ensuremath{\Bu}\xspace}

\begin{document}

\preprint{\vbox{
    \hbox{Belle Preprint 2022-30}
    \hbox{KEK Preprint 2022-41}
}}

\title{Search for the lepton flavour violating decays $B^{+} \to K^{+} \tau^\pm \ell^\mp$~($\ell = e, \mu$) at Belle}
\include{pub635-orcid_2}

\begin{abstract}
We present a search for the lepton-flavour-violating decays $B^+ \to K^+ \tau^\pm \ell^\mp$, with~$\ell = (e, \mu)$, using the full data sample of $772 \times 10^6$ $B\overline{B}$ pairs recorded by the Belle detector at the KEKB asymmetric-energy $e^+ e^-$ collider. We use events in which one $B$ meson is fully reconstructed in a hadronic decay mode. We find no evidence for $B^\pm \to K^\pm \tau \ell$ decays and set upper limits on their branching fractions at the 90\% confidence level in the $(1$--$3) \times 10^{-5}$ range. 
The obtained limits are the world's best results.
\end{abstract}

\maketitle

Recently, there has been a resurgence of interest in the study of leptoquark fields in light of discrepancies in semi-leptonic $B$-decays~\cite{nature}, collectively known as $B$-physics anomalies, which challenge the assumed Lepton Flavour Universality (LFU) of fundamental interactions. These measurements have been obtained studying two different quark transitions: $b \to c \tau \nu$ and $ b \to s \ell \ell$, where~$\ell = (e, \mu)$.
If confirmed by further measurements, this would be clear evidence of New Physics in which new heavy particles couple preferentially to second and third generation leptons. 
Many extensions of the Standard Model (SM) that include violation of LFU predict Lepton Flavour Violating (LFV) processes in hadron decays with charged leptons in the final state~\cite{glashow}.
In particular, the vector leptoquark $U_1$ with SM quantum numbers $(\mathbf{3}, \mathbf{1})_{2/3}$ has been identified as the only single-mediator solution~\cite{Angelescu:2021lln}. In the minimal scenario, $U_1$ provides an interesting prediction, i.e.\ lower bounds on the lepton flavour violating $b \to s \tau^{\mp} \mu^{\pm}$ decay modes, for example ${\cal B}(B \to K\tau\mu) > 0.7 \times 10^{-7}$.
The branching fractions for the two $\ell \tau$ charge combinations are not in general the same, as they depend on the details of the physics mechanism producing the decay.

Upper limits on the branching fractions for $B^+ \to K^+ \tau^\pm \ell^\mp$ decays have been previously set at the 90\% confidence level (C.L.) using hadronic $B$-tagging by the BaBar collaboration between $1.5 \times 10^{-5}$ and $4.5 \times 10^{-5}$~\cite{babar}; the LHCb~collaboration has studied a single mode, using $B^+$ mesons from $B_{s2}^{*0} \to B^+ K^-$ decays, setting a limit ${\cal B}(B^+ \to K^+\tau^+\mu^-) < 3.9 \times 10^{-5}$ at the 90\% C.L.~\cite{LHCb:2020khb}.

In this Letter, we report a search for $B^+ \to K^+ \tau^{\mp}\ell^{\pm}$ decays  using the full Belle data sample recorded at the $\Upsilon(4S)$ resonance. 
The inclusion of the charge-conjugate decay mode is implied.
This is the first such search from Belle.
 


The analysis is based on the full data sample of $772 \times 10^6$ $B\overline{B}$ pairs collected with the Belle detector~\cite{Belle:2000cnh} at the KEKB asymmetric-energy $e^+e^-$ collider~\cite{Kurokawa:2001nw}. 
The Belle detector is a large-solid-angle spectrometer, which includes a silicon vertex detector (SVD), a 50-layer central drift chamber (CDC), an array of aerogel threshold Cherenkov counters (ACC), time-of-flight scintillation counters (TOF), and an electromagnetic calorimeter (ECL) comprised of CsI(Tl) crystals located inside a superconducting solenoid coil that provides a 1.5~T magnetic field. 
An iron flux return located outside the coil is instrumented to detect $K^{0}_{L}$ mesons and identify muons.


The analysis procedure is developed using Monte Carlo (MC) simulation based on events generated with \textsc{EvtGen}~\cite{evtgen}, which includes final-state radiation (FSR) effects simulated by PHOTOS~\cite{photos}, and the detector response is simulated by \textsc{GEANT3}~\cite{geant}. 
The $B^+ \to K^+ \tau^\pm \ell^\mp$ decays are generated using a uniform three-body phase space model (PHSP); we also consider variations in the linear combinations of the relevant operators for the $b \rightarrow s\tau\ell$ transitions: $\mathcal{O}_{i}$ and the relative Wilson coefficients $C_i^{\tau\ell}$, where $i=9,10,S,P$~\cite{Becirevic:1602}.

In each event, we require a fully reconstructed hadronic $B^{\pm}$ decay, which we refer to as the tagged $B$ meson candidate or $B_{\rm tag}$.
This is done using the Full Event Interpretation (FEI) algorithm~\cite{fei}, a machine-learning algorithm developed for $B$-tagged analyses at Belle and Belle~II.
It supports both hadronic and semileptonic tagging, reconstructing $B$ mesons across more than 4000 individual decay chains. 
The training is performed in a hierarchical manner: final-state particles are first reconstructed from detector information, then unstable particles (e.g.\ $D$, $D^*$) are built up from these particles, and then reconstruction of $B$-mesons is performed last.
For each $B_{\rm tag}$ candidate reconstructed by the FEI, a value of the final multi-variate classifier output, $\Sigma_{\rm FEI}$, is assigned. 
$\Sigma_{\rm FEI}$ is distributed between zero and one, representing candidates identified as being background-like and signal-like, respectively.

For the hadronic FEI, the minimal number of tracks per event satisfying certain quality criteria is set to three, as the vast majority of $B$-meson chains include at least three charged particles, and such a criterion is useful for suppressing background from non-$B\overline{B}$ events.
Requirements are placed on the impact parameters to ensure close proximity to the interaction point~(IP), less than $0.5$~cm in the transverse plane and less than $2.0$~cm along the $z$ axis (parallel to the $e^{+}$ beam). 
ECL clusters that are used for $\gamma$ reconstruction are required to satisfy a region-dependent energy threshold criterion. All the intermediate states ($\pi^0, J/\psi,K_S^0$ and $D^{(*)}$ mesons)  must pass loose cuts on the reconstructed invariant mass and only the best candidates in terms of $\Sigma_{\rm FEI}$ are kept.
The FEI results in many $B_{\rm tag}$ candidates per event. 
The number of these candidates is reduced with selections on the beam-energy-constrained mass $\Mbc = \sqrt{(E^{*}_{\rm beam}/c^{2})^{2} - (p^{*}_{{B}_{\rm tag}}/c)^{2}}$, and the energy difference $\Delta E =  E^{*}_{B_{\rm tag}} - E^{*}_{\rm beam}$, where $E^{*}_{\rm beam}$ is the beam energy, and  $E^{*}_{B_{\rm tag}}$ and $p^{*}_{B_{\rm tag}}$ are the energy and momentum of the $B_{\rm tag}$ candidate in the center of mass (c.m.) rest frame, respectively.
The criteria applied are $\Mbc > 5.27$~GeV/$c^2$ and $|\Delta E| < 0.1$~GeV.
Finally, the candidate with the highest $B_{\rm tag}$ classifier output, $\Sigma_{\rm FEI}$, is selected and a loose requirement $\Sigma_{\rm FEI}$ $> 0.001$, provides further background rejection with little signal loss.


We then search for the signal $B \to K \tau \ell$ decay in the rest of the event, which we refer to as the signal $B$ meson candidate or $B_{\rm sig}$. 
The notation $B \to K \tau \ell$ refers to one of the following four final states that we consider, where in addition to the kaon of opposite charge to $B_{\rm tag}$ we associate the primary lepton, $\mu$ or $e$: $B^+ \to K^+ \tau^+ \mu^-$ and $B^+ \to K^+ \tau^+ e^-$ defined as $OS_{\mu, e}$ modes because the kaon and the primary lepton have opposite charge, and $B^+ \to K^+ \tau^- \mu^+$ and $B^+ \to K^+ \tau^- e^+$, defined as $SS_{\mu, e}$ modes. In all cases, we require that the $\tau$ decays to $\tau \to e \nu \overline{\nu}$, $\tau \to \mu \nu \overline{\nu}$, or $\tau \to \pi \nu$. The combined branching fraction for these decays is 46\%~\cite{ParticleDataGroup:2022pth}. The $\tau \to \rho \nu$ mode, despite not being explicitly reconstructed, significantly contributes to the $\tau \to \pi \nu$ candidates -- by roughly one half -- because of its large branching fraction ($\sim25\%$).

We reconstruct $B^+ \to K \tau^{\pm} \ell^{\mp}$ decays by selecting three charged particles that originate from the vicinity of the~IP and are not associated with the $B_{\rm tag}$.
We require impact parameters less than $1.0$~cm in the transverse plane and less than $4.0$~cm along the $z$ axis. 
To reduce backgrounds from low-momentum particles, we require that tracks have a minimum transverse momentum of 100~MeV/$c$.
From the list of selected tracks, we identify $K^{+}$ candidates using a likelihood ratio ${\cal R}^{}_{K/\pi} = {\cal L}^{}_K / ({\cal L}^{}_K + {\cal L}^{}_\pi )$, where ${\cal L}^{}_{K}$ and ${\cal L}^{}_{\pi}$ are the likelihoods for charged kaons and pions, respectively, calculated based on the number of photoelectrons in the ACC, the specific ionization in the CDC, and the time of flight as determined from the TOF. 
We select kaons by requiring ${\cal R}^{}_{K/\pi} > 0.6$, which has a kaon identification efficiency of 83\% and a pion misidentification rate of 5\%. 
Similarly, we select pions by requiring ${\cal R}^{}_{\pi/K} > 0.6$, which has a pion identification efficiency of 84\% and a kaon misidentification rate of 6\%.

Muon candidates are identified based on information from the KLM. We require that candidates have a momentum greater than 0.8~GeV/$c$~(enabling them to sufficiently penetrate KLM), and a penetration depth and degree of transverse scattering consistent with those of a muon~\cite{muid}. 
The latter information is used to calculate a normalized muon likelihood ratio ${\cal R}_{\mu} = {\cal L}_{\mu} / ({\cal L}_{\mu} + {\cal L}^{}_K + {\cal L}^{}_\pi)$, where ${\cal L}_{\mu}$ is the likelihood for muons, for which we require ${\cal R}_{\mu} > 0.9$. 
For this requirement, the average muon detection efficiency is  89\%, with a pion misidentification rate of 1.5\%~\cite{pid}.

Electron candidates are required to have a momentum greater than 0.5~\gevc~and are identified using the ratio of ECL cluster energy to the CDC track momentum, the shower shape in the ECL, the matching of the track with the ECL cluster, the specific ionization in the CDC, and the number of photoelectrons in the ACC.
This information is used to calculate a normalized electron likelihood ratio ${\cal R}^{}_{e} = {\cal L}_{e} / ({\cal L}_{e} + {\cal L}_{\rm hadrons})$, where ${\cal L}_{\rm hadrons}$ is a product of hadron likelihoods, for which we require ${\cal R}^{}_{e} >~0.9$.
This requirement has an efficiency of 92\% and a pion misidentification rate below 1\%~\cite{eid}. 

After selecting one charged kaon, one prompt lepton (electron or muon) and the $\tau$ daughter (electron, muon or pion) with the appropriate charge combination, we require that there are no other tracks than the ones associated to $B_{\rm tag}$ or $B_{\rm sig}$.
The charged kaon and the prompt lepton are uniquely determined to minimize $\chi^{2}$ of the $B_{\rm sig}$ vertex fit for the prompt tracks. In case there are two possibilities in $\tau$ daughter particle identification, $\tau$ leptonic decay has priority.
Unlike other $B$ decays involving $\tau$'s (e.g. $B \to \tau \nu$, $B \to D^* \tau \nu$), the $B \to K \tau \ell$ channel has the unique property of having the one (or two) neutrino(s) coming only from the $\tau$ itself, allowing the signal yield to be extracted using the recoil mass, $M_{\rm recoil}$, which should peak at the mass of the $\tau$ lepton.
Such variable is easily obtained at $B$-factories, because of the known initial kinematics and the full reconstruction of the other $B$ in the event.
In fact, if we consider the $B_{\rm sig}$, the 4-momentum of the $\tau$ can be written as:
\begin{linenomath}
\begin{equation}
\label{eq:momconv}
p_{\tau} = p_{B_{\rm sig}}-p_{K}-p_{\ell} \\
\end{equation}
\end{linenomath}
where $p_{B_{\rm sig}}$ is not known a priori. In the frame where the \FourS resonance is at rest, 
the two $B$ mesons are back to back, hence:
\begin{linenomath}
\begin{align}
\mathbf{p}^{*}_{B_{\rm tag}} &= - \mathbf{p}^{*}_{B_{\rm sig}}
\label{eq:momcon}
\end{align}
\end{linenomath}
furthermore, the two $B$'s have the same energy, which is half the energy $\sqrt{s}$ of the \FourS:
\begin{linenomath}
\begin{align}
{E}^{*}_{B_{\rm tag}} &= {E}^{*}_{B_{\rm sig}} = \frac{\sqrt{s}}{2}.
\label{eq:encon}
\end{align}
\end{linenomath}
In order to obtain the best resolution on the $B$ variables, we replace ${E}^{*}_{B_{\rm tag}}$ with $E_{\rm beam}^{*}$, but use the reconstructed ${p}^{*}_{B_{\rm tag}}$ rather than the average value $p_{\rm beam}^{*}=\sqrt{E_{\rm beam}^{*2}/c^2-m_{B}^2c^2}$.
Using the condition~(\ref{eq:momcon}) and the substitution $E_{B}^{*}=E^*_{\rm beam}$ in equation~(\ref{eq:momconv}), we obtain:
\begin{widetext}
\begin{equation}
\begin{cases}
\label{eq:mrecformula}
\mathbf{p}_{\tau}^{*} &= -\mathbf{p}_{B_{\rm tag}}^{*}-\mathbf{p}_{K}^{*}-\mathbf{p}_{\ell}^{*} \\
{E}_{\tau} &= {E}_{\rm beam}^{*}-E_{K}^{*}-E_{\ell}^{*} \\
\end{cases} 
\implies 
M_{\rm recoil}^2 = m_{\tau}^2 = m_{B}^2 + 
m_{{K}\ell}^2- 2(E^*_{\rm beam} E^*_{{K}\ell}/c^4+
         p^*_{B_{\rm tag}}p^*_{{K}\ell}\cos\theta/c^2 )
\end{equation}
\end{widetext}
where $\theta$ is the angle between $\mathbf{p}^*_{B_{\rm tag}}$ and $\mathbf{p}^*_{{K}\ell}$.

The main source of background consists of Cabibbo-favoured transitions from $B^+B^-$ events. 
For the $OS$ configurations, where the primary lepton charge is opposite to the $B_{\rm sig}$ charge, the dominant background comes from semileptonic $D$ decays: $\Bp \to \overline{D}^{0}(\to K^+\ell^-\overline{\nu}_{\ell})X^+$.
On the other hand, for the $SS$ configurations the primary lepton and the $B_{\rm sig}$ have the same charge and the semileptonic $B^+$ decays like $\Bp \to \overline{D}^{0}(\to K^+X^-)X\ell^+\nu_{\ell}$ provide the three charged particles for the $B_{\rm sig}$ candidates. Events compatible with a $B^+_{\rm sig}\to \overline{D}^{0}(\to{ K^+\pi^-)X^+}$ decay are rejected by vetoing candidates in the range $1.81$~GeV$/c^2<m_{K^+t^-}<1.91$~GeV$/c^2$, where $t$ denotes the primary lepton or the track from the $\tau$ in the $OS$ or $SS$ case, respectively. 
In the first case, only the $K\tau\mu$ modes show such a $D^0$ component because of the larger probability to identify a pion as a muon rather than an electron. 
In the $SS$ case, the $D^0$ peak is much more prominent and relates to the $\tau\to\pi$ mode and is independent of the flavour of the primary lepton.

We further improve the signal selection using a Boosted Decision Tree (BDT) classification.
Two classifiers are trained for the background suppression. The first one is optimised to reduce the $B\overline{B}$ events and uses as inputs some kinematic information as well as the topology of the $B_{\rm sig}$ and information on the rest of the event (the set of ECL clusters that are not used for the $B_{\rm sig}$ and $B_{\rm tag}$ reconstruction). In particular we use: the invariant mass $m_{K^+t^-}$, which helps in suppressing the combinatorial background from charm decays, the number of ECL clusters that are not associated with the reconstructed event and the sum of their energies, the extended Fox-Wolfram moments~\cite{KFSW}, the distance from the IP of the signal vertex and the distance between the primary kaon and each of the other two signal tracks. 
For each mode only the ten most important variables are kept for the final training, the metrics being the information gain provided by each feature in all the decision trees used for the classifier.
The threshold $t$ on the BDT response is optimized using a figure of merit~\cite{punzi}, defined as 
\begin{linenomath}
\begin{equation}
\label{eq:punziFOM}
    \mathcal{F}(t) = \frac{\epsilon (t)}{\frac{3}{2}+\sqrt{N_{\rm bkg}(t)}},
\end{equation}
\end{linenomath}
where $\epsilon (t)$ is the efficiency for the cut $t$, $N_{\rm bkg}$ represents the number of background events surviving the cut $t$ in the signal region defined as $1.68$~GeV/$c^{2}$ $< M_{\rm recoil} < 1.87$~GeV/$c^{2}$ which contains $\sim80\%$ of the signal events. 
The MC sample used to estimate the background corresponds to a luminosity of twice that of data. 
After the cut on the first BDT output, a large fraction of the surviving background is coming from $q\overline{q}$ ($q=u,d,s,c$) events; for this reason a second BDT classifier is trained on these events. The input variables to suppress the continuum background are: event-shape variables such as $R_2$ and the CLEO cones~\cite{cleoC} and the angle $\theta_{\rm T}$ between the thrust axes calculated from final-state particles for the $B_{\rm tag}$ and for the rest of the event in the c.m. frame.

\begin{figure}[h]
\includegraphics[scale=0.42]{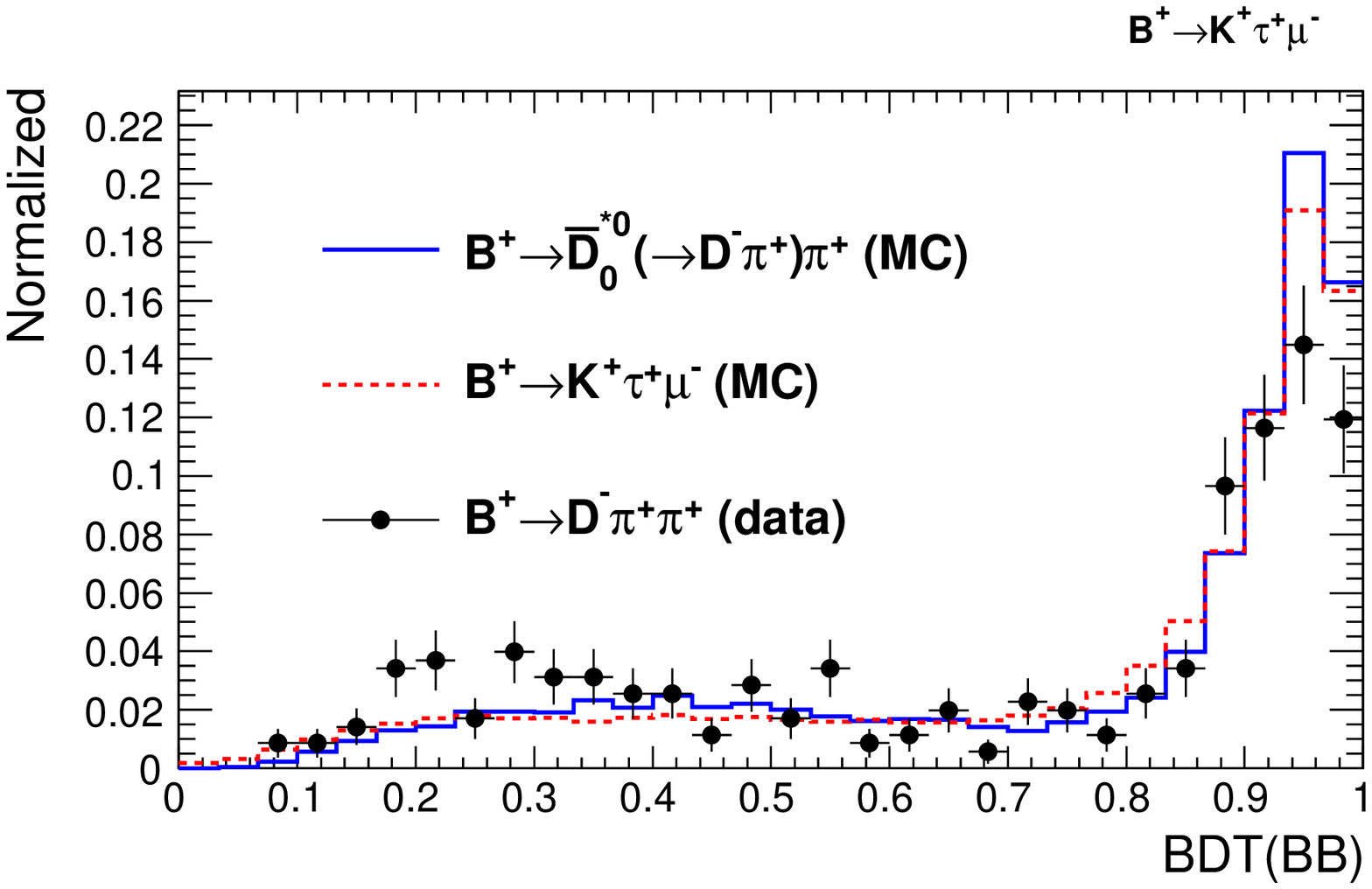}\\
\includegraphics[scale=0.42]{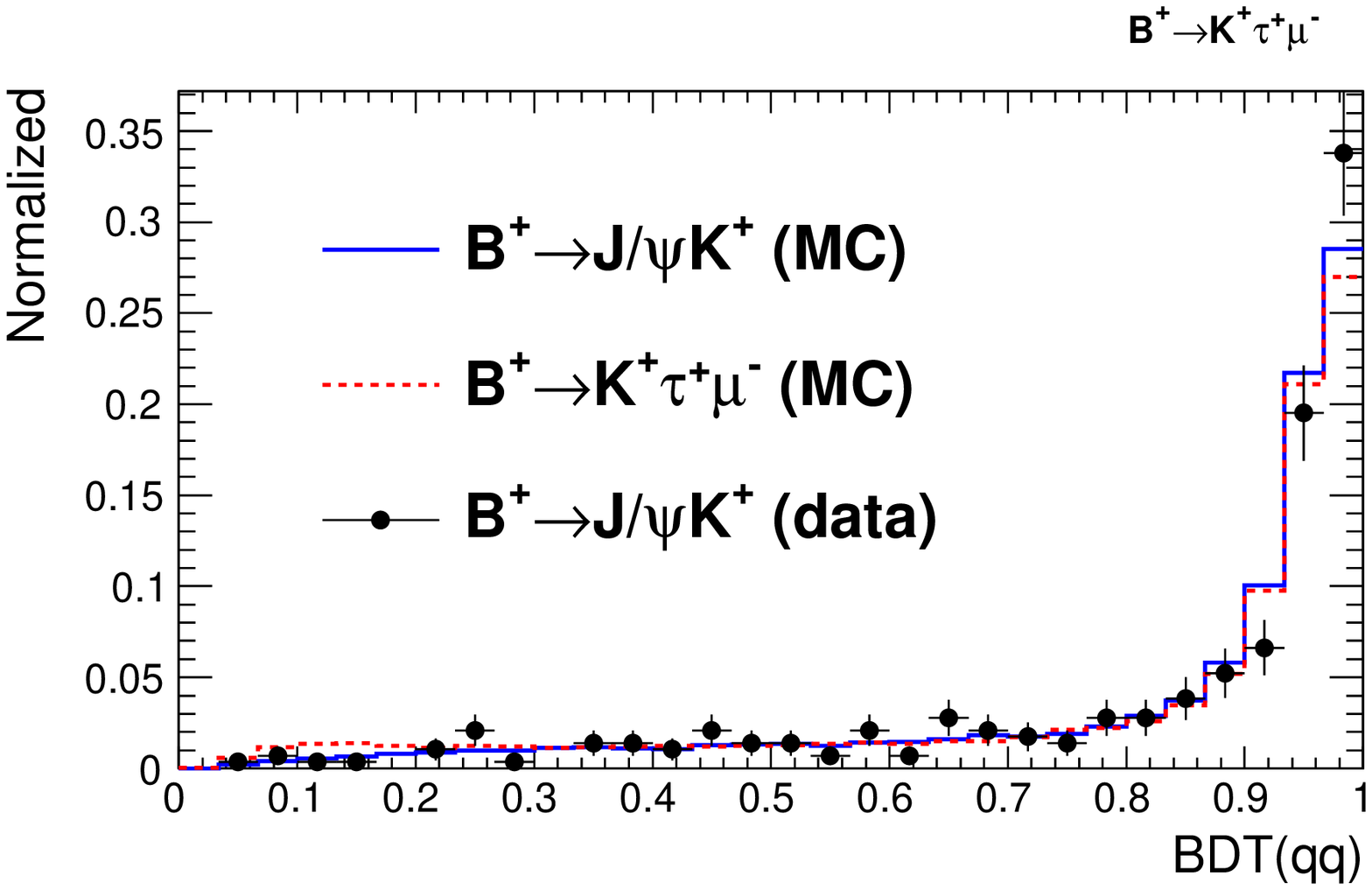}
\caption{The BDT response of the $B\overline{B}$ suppression for the $B^+ \to D^-\pi^+\pi^+$ mode (top), and the $q\overline{q}$ suppression for the $B^+ \to J/\psi K^+$ mode (bottom) for $B^+ \to K^+ \tau^+ \mu^-$ case.  \label{fig:bdts}}
\end{figure}

We use control samples in order to evaluate systematic uncertainties related to data/MC discrepancies and to calibrate the signal shape PDF as it is fixed from MC simulation. The first control sample consists of $B^+ \to D^-\pi^+\pi^+$ events, generated in MC as the result of two-body $B^+\to\overline{D}^{**0}(\to D^-\pi^+)\pi^+$ decays, with $\overline{D}^{**0}=\{\overline{D}_0^{*0},\overline{D}_2^{*0}\}$ according to Refs.~\cite{BaBar:2009pnd,Belle:2003nsh}. This channel has similar topology to our signal as the $D$ can be treated as the $\tau$, allowing for a comparison of the performance of the first BDT classifier between data and MC (Fig.~\ref{fig:bdts}(top) for $OS_\mu$). 
For the calibration of the efficiency of $q\overline{q}$ suppression a second control sample $B \to J/\psi K$ is used because of the similar final state while no usage of the $B_{\rm sig}$ topology is required  (Fig.~\ref{fig:bdts}(bottom) for $OS_\mu$). 

The signal yields for $B \to K\tau\ell$ decays are obtained by performing unbinned extended maximum-likelihood fits to the $M_{\rm recoil}$ distributions.
The PDF used to model reconstructed signal decays consists of the sum of a reversed Crystal Ball function to model the main peak and the high-side power-law tail, and a broad Gaussian, with the same mean parameter, to describe the candidates with worse resolution due to imperfect $B_{\rm tag}$ reconstruction.
The background events have a smooth shape in the $M_{\rm recoil}$ signal region, and are described by a $2$nd-order Chebychev polynomial. 
The yields are floated, as well as the background shape parameters while the parameters describing the signal PDF are fixed from the MC simulation.
We apply corrections to these parameters to account for small differences between MC simulation and data. 
These correction factors are obtained from the $B^+ \to \overline{D}{}^{(*)0} \pi^+$ control samples where $M_{\rm recoil}$ is calculated from the pion from the $B^+$ decay and the $B_{\rm tag}$.
We validate our fitting procedure and check for fit bias using MC simulation. 
We generate large ensembles of simulated experiments in which the $M_{\rm recoil}$ distributions are generated from the PDFs used for fitting.

\begin{figure*}
\includegraphics[scale=0.40]{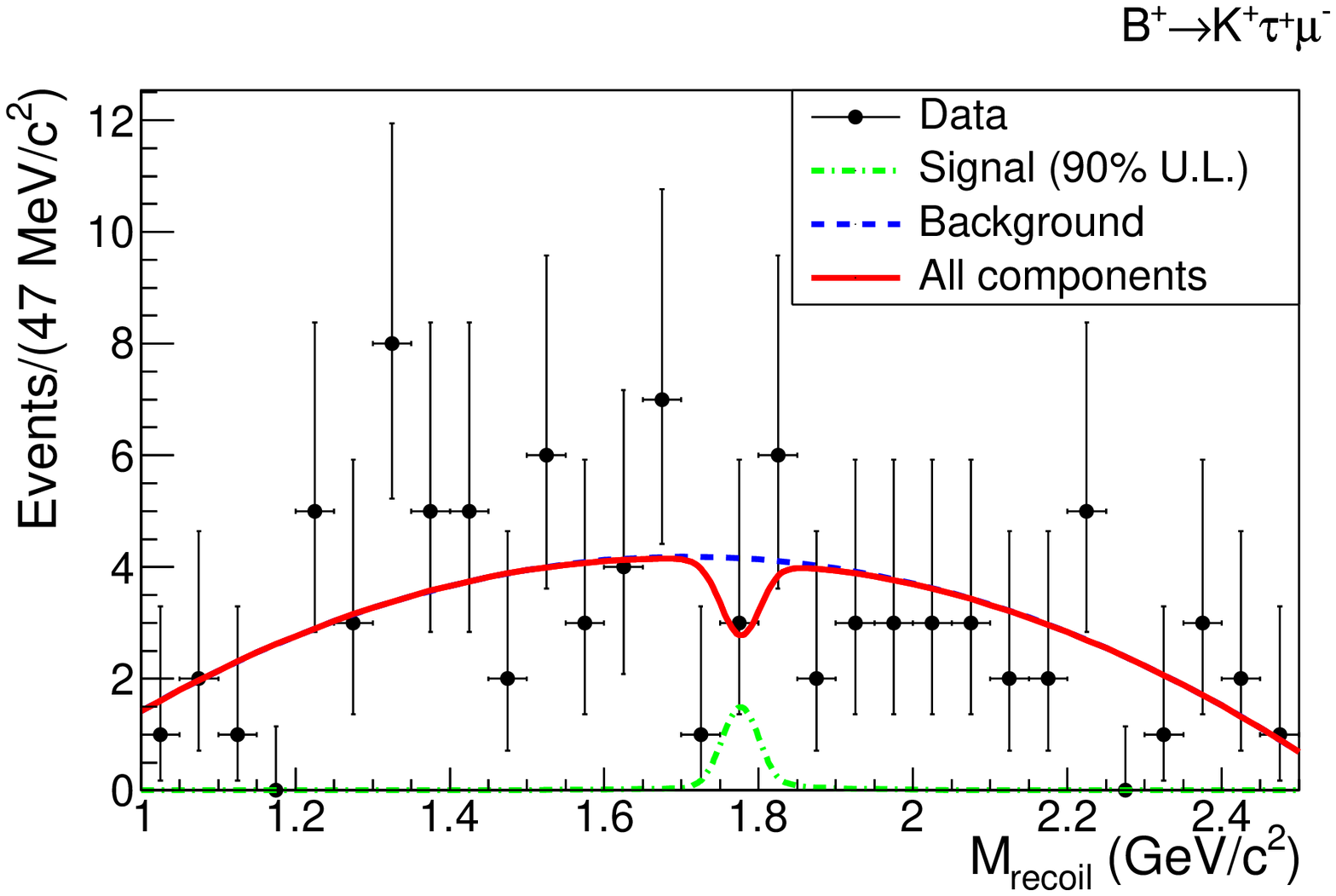}
\includegraphics[scale=0.40]{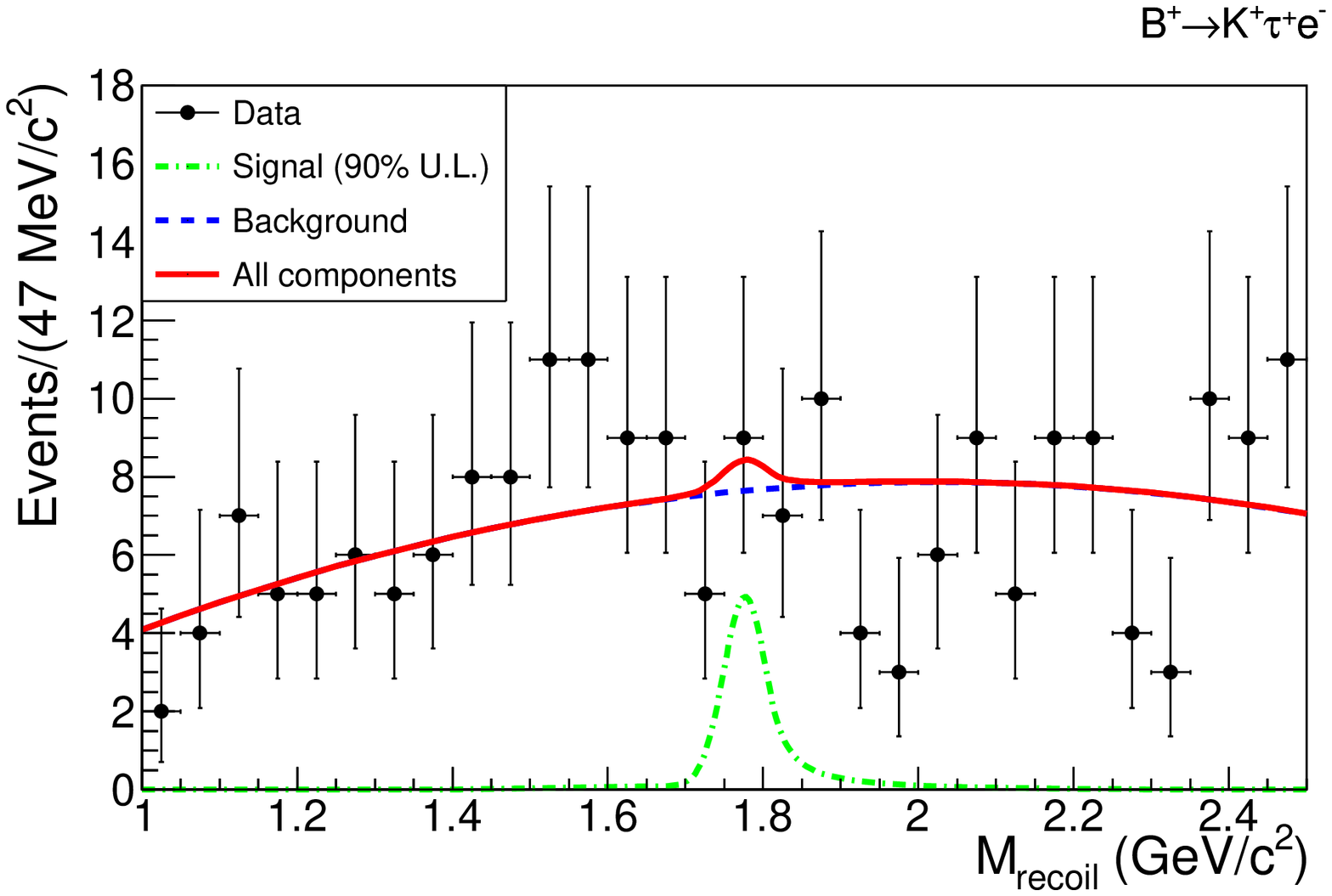}\\
\includegraphics[scale=0.40]{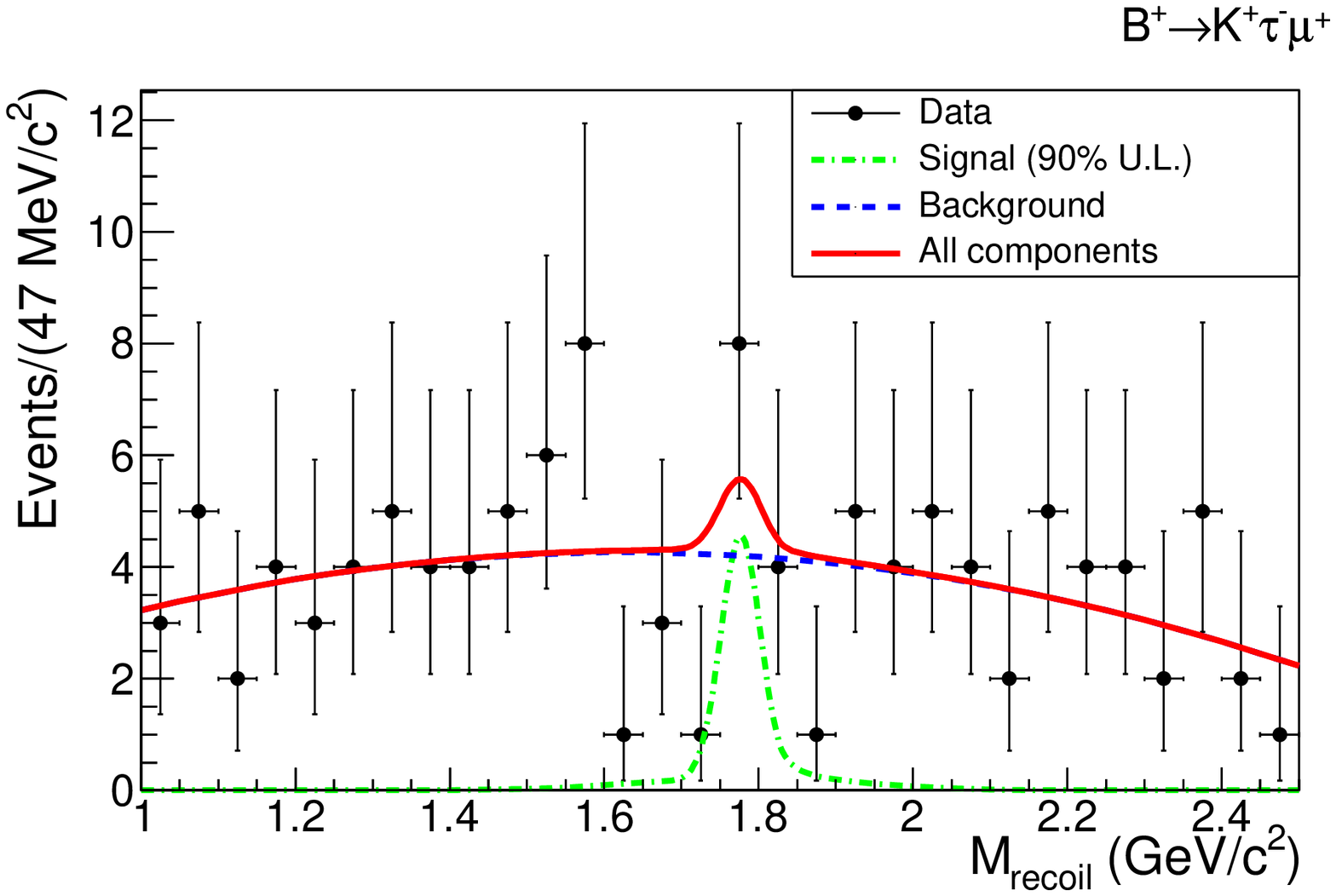}
\includegraphics[scale=0.40]{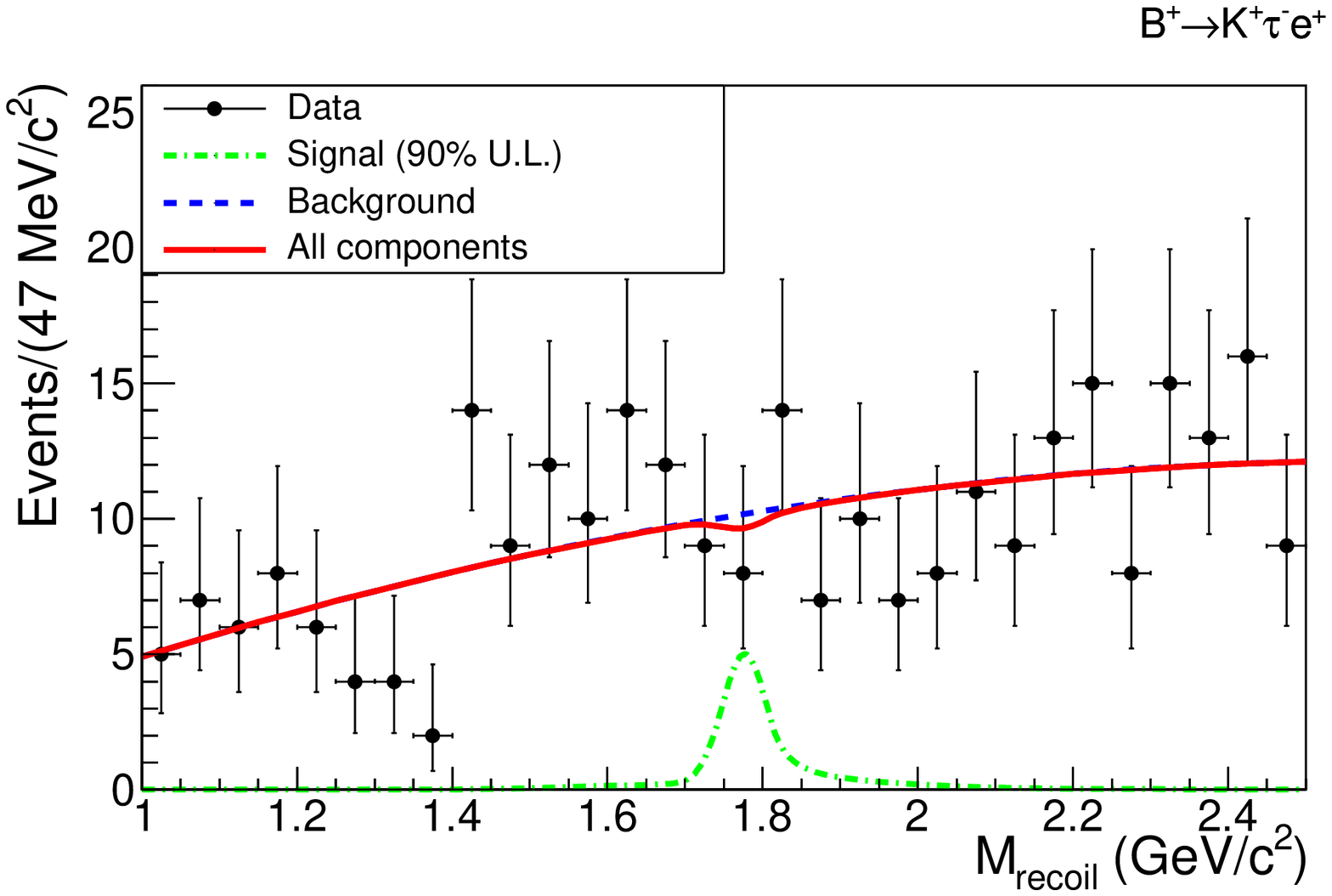}
\caption{Observed $M_{\rm recoil}$ distributions for the four $B \to K \tau \ell$ modes, along with projections of the fit result. The black dots show the data, the dashed blue curve shows the background component, and the solid red curve shows the overall fit result. The dash-dotted green curve shows the signal PDF, with a normalization corresponding to the 90\% C.L. upper limit. \label{fig:btoktaul}}
\end{figure*}

The $M_{\rm recoil}$ distributions for LFV $B \to K \tau \ell$ decays along with projections of the fit result are shown in Fig.~\ref{fig:btoktaul}. 
The fitted signal yields listed in Table~\ref{LFV_table} are consistent with zero for all four modes.

We calculate the upper limit (UL) for these modes at the $90\%$~C.L. using a frequentist method. In this method, for different numbers of signal events $N_{\rm{sig}} (\rm{gen})$, we generate 10000 pseudo experiments with signal and background PDFs as obtained in the nominal data fit, with each set of events being statistically equivalent to our data sample of 711~fb$^{-1}$. We fit all these simulated data sets, and, for each value of $N_{\rm{sig}}(\rm{gen)}$, we calculate the fraction of MC experiments that have $N_{\rm sig} \leq N_{\rm{sig}}(\rm{data})$. The $90\%$ C.L. upper limit is taken to be the value of $N_{\rm{sig}}(\rm{gen})$ (called here $N_{\rm sig}^{\rm{UL}}$) for which $10\%$ of the experiments have $N_{\rm{sig}} \leq N_{\rm{sig}}(\rm{data})$. The upper limit on the branching fraction is then derived using the formula:
\begin{center}
$\cal B^{\rm{UL}}$ $= \dfrac{N_{\rm sig}^{\rm{UL}}}{N_{B\overline{B}} \times 2 \times f^{+-} \times \varepsilon}$,
\end{center} 
where $N_{B\overline{B}}$ is the number of $B\overline{B}$ pairs = $(772 \pm 11) \times 10^{6}$, $f^{+-}$ is the branching fraction ${\cal B}(\Upsilon(4S) \to B^{+}B^{-})$ for charged $B$ decays (using $0.514 \pm 0.006$~\cite{ParticleDataGroup:2022pth}), and $\varepsilon$ is the signal reconstruction efficiency. By default $\varepsilon$ is obtained with signal PHSP MC samples~\cite{efficiency}, while we also consider a NP model with a combination of the effective operators $\mathcal{O}_{S,P}$ by reweighting the $q^2 = m_{\tau \ell}^2$ distribution which gives the smallest efficiency.
The systematic uncertainty in $\cal B^{\rm{UL}}$ is included by smearing the $N_{\rm sig}$ distribution obtained from the MC fits with the fractional systematic uncertainty. The results are listed in Table~\ref{LFV_table}.

\begin{table}[h]
\begin{center}
\caption{Efficiencies, fit yields, and branching fraction upper limits at the 90\% C.L. for PHSP (and NP) case. }
\label{LFV_table}
\begin{tabular}{l c c c c c}
Mode & $\varepsilon$ (\%) & $\varepsilon^{\rm NP}$ (\%) & $N_{\rm sig}$ & 
$\cal B^{\rm{UL}}$ $(10^{-5})$ \\ 
\hline
$B^{+} \rightarrow K^{+}\tau^{+}\mu^{-}$ & 0.064 & 0.058 & $-2.1 \pm 2.9$  & 0.59 (0.65)\\
$B^{+} \rightarrow K^{+}\tau^{+}e^{-}$ & 0.084 & 0.074 & $\;\;\;1.5 \pm 5.5$  & 1.51 (1.71)\\
$B^{+} \rightarrow K^{+}\tau^{-}\mu^{+}$ & 0.046 & 0.038 & $\;\;\;2.3 \pm 4.1$  & 2.45 (2.97) \\
$B^{+} \rightarrow K^{+}\tau^{-}e^{+}$ & 0.079 & 0.058 & $-1.1 \pm 7.4$  & 1.53 (2.08) \\
\hline
\end{tabular}
\end{center}
\end{table}

\begin{table}[h]
\centering
\caption{Contributions to the systematic uncertainties of the measurements.}

\begin{tabular}{l
      S[table-format=-3.2]
      S[table-format=-3.2]
      S[table-format=-3.2]
      S[table-format=-3.2]}
Source &   {$K^+\tau^+ \mu^-$} &  {$K^+\tau^+ e^-$} &  {$K^+\tau^- \mu^+$} &  {$K^+\tau^- e^+$} \\ 
\hline
Additive (events) \\
\hline
PDF shape (mean) & 0.09 & 0.01 & 0.08 & 0.08 \\
PDF shape (width) & 0.02 & 0.08 & 0.04 & 0.07\\
PDF shape ($f_{\rm sig}$) & 0.28 & 0.16 & 0.11 & 0.16 \\
Linearity     &  0.03 & 0.04  & 0.02 & 0.04 \\
\hline
Total & 0.30 & 0.18 & 0.14 & 0.20 \\
\hline
Multiplicative (\%) \\
\hline
$B_{\rm tag}$ calibration & 5.9 & 5.9 & 5.9 & 5.9 \\
Track reconstruction  & 1.1 & 1.1 & 1.1 & 1.1 \\
Kaon id.     & 1.3 & 1.4 & 1.3 & 1.3 \\
Lepton id.   & 0.3 & 0.4 & 0.3 & 0.4 \\
$\tau$ daughter id.  & 0.7 & 0.7 & 0.6 & 0.6 \\
MC statistics     &  1.0 & 1.5 & 1.2 & 1.0 \\
Number of $B \overline{B}$ pairs & 1.4 & 1.4 & 1.4 & 1.4 \\
BDT $B\overline{B}$ selection & 10.6 & 10.0 & 12.7 & 12.6 \\
BDT $q\overline{q}$ selection & 8.8 & 8.6 & 9.2 & 6.6 \\
$f^{+-}$     & 1.2 & 1.2 & 1.2 & 1.2 \\
\hline
 Total & 15.3 & 14.8 & 17.0 & 15.7 \\
\hline 
\end{tabular}
\label{tab:tablebf}
\end{table}

The systematic uncertainties in our measurements are listed in Table~\ref{tab:tablebf},
where additive uncertainties arise from the signal yield, while multiplicative uncertainties are from the efficiency.
Uncertainties in the shape of the PDFs used for the signal are evaluated by varying all fixed parameters by $\pm 1 \sigma$, including the correction factors to the shapes obtained from the $B^+ \to \overline{D}^{(*)0} \pi^+$ control samples, and varying the fraction of the Gaussian ($f_{\rm sig}$) by 10\%.
The resulting change in the signal yield is taken as the systematic uncertainty.
The reconstruction efficiency for $B_{\rm tag}$ evaluated via MC simulation, is corrected to account for differences between MC and data in the branching fractions and models used for hadronic $B$ decays.
This correction is evaluated by comparing the number of events containing both a $B_{\rm tag}$ and a semileptonic $B \to D\ell \nu$~\cite{Gelb}.
The resulting correction factor is $85 \pm 5$\% and the uncertainty in this value is taken as a systematic uncertainty.

The systematic uncertainty due to the charged track reconstruction is evaluated using $D^{*+} \to D^0 \pi^+$ with $D^0 \to K_S^0 \pi^+\pi^-$, resulting in an uncertainty of 0.35\% per track. Uncertainties due to $K^+$ and $\pi^{+}$ (for $\tau\rightarrow\pi\nu$ mode) identification is 1.3\%, as measured with a $D^{*+}\to
D^0(K^-\pi^+)\pi^+$ sample. The uncertainty due to lepton identification is evaluated using $J/\psi \to \ell^+ \ell^-$ events, resulting in an uncertainty of 0.3\% for muons and 0.4\% for electrons.
The systematic uncertainty arising from the number of $B \overline{B}$ pairs is 1.4\%.
We compare the efficiency of the BDT selection between data and MC samples with the control channel $B^+ \to D^-\pi^+\pi^+$ for $B\overline{B}$ suppression and $B^+ \to J/\psi K^+$ for continuum suppression, the differences between data and MC simulation are assigned as a systematic uncertainty. 
We use a systematic uncertainty of 1.2\% in the fraction $f^{+-}$~\cite{ParticleDataGroup:2022pth}.

We have searched for the lepton-flavour-violating decays $B^+ \to K^+ \tau^{\pm}\ell^{\mp}$ using the full Belle data set. 
We find no evidence for these decays and set the following upper limits on the branching fractions at the 90\% C.L.: 
\begin{linenomath}
\begin{align}
\begin{split}
  &{\cal B}(B^{+} \rightarrow K^{+}\tau^{+}\mu^{-}) < 0.59 \times 10^{-5} \\
  &{\cal B}(B^{+} \rightarrow K^{+}\tau^{+}e^{-}) < 1.51 \times 10^{-5} \\
  &{\cal B}(B^{+} \rightarrow K^{+}\tau^{-}\mu^{+}) < 2.45 \times 10^{-5} \\
  &{\cal B}(B^{+} \rightarrow K^{+}\tau^{-}e^{+}) < 1.53 \times 10^{-5}
\end{split}
\end{align}
\end{linenomath}
Our results are the most stringent limits to date.\\

This work, based on data collected using the Belle detector, which was
operated until June 2010, was supported by 
the Ministry of Education, Culture, Sports, Science, and
Technology (MEXT) of Japan, the Japan Society for the 
Promotion of Science (JSPS), and the Tau-Lepton Physics 
Research Center of Nagoya University; 
the Australian Research Council including grants
DP180102629, 
DP170102389, 
DP170102204, 
DE220100462, 
DP150103061, 
FT130100303; 
Austrian Federal Ministry of Education, Science and Research (FWF) and
FWF Austrian Science Fund No.~P~31361-N36;
the National Natural Science Foundation of China under Contracts
No.~11675166,  
No.~11705209;  
No.~11975076;  
No.~12135005;  
No.~12175041;  
No.~12161141008; 
Key Research Program of Frontier Sciences, Chinese Academy of Sciences (CAS), Grant No.~QYZDJ-SSW-SLH011; 
Project ZR2022JQ02 supported by Shandong Provincial Natural Science Foundation;
the Ministry of Education, Youth and Sports of the Czech
Republic under Contract No.~LTT17020;
the Czech Science Foundation Grant No. 22-18469S;
Horizon 2020 ERC Advanced Grant No.~884719 and ERC Starting Grant No.~947006 ``InterLeptons'' (European Union);
the Carl Zeiss Foundation, the Deutsche Forschungsgemeinschaft, the
Excellence Cluster Universe, and the VolkswagenStiftung;
the Department of Atomic Energy (Project Identification No. RTI 4002) and the Department of Science and Technology of India; 
the Istituto Nazionale di Fisica Nucleare of Italy; 
National Research Foundation (NRF) of Korea Grant
Nos.~2016R1\-D1A1B\-02012900, 2018R1\-A2B\-3003643,
2018R1\-A6A1A\-06024970, RS\-2022\-00197659,
2019R1\-I1A3A\-01058933, 2021R1\-A6A1A\-03043957,
2021R1\-F1A\-1060423, 2021R1\-F1A\-1064008, 2022R1\-A2C\-1003993;
Radiation Science Research Institute, Foreign Large-size Research Facility Application Supporting project, the Global Science Experimental Data Hub Center of the Korea Institute of Science and Technology Information and KREONET/GLORIAD;
the Polish Ministry of Science and Higher Education and 
the National Science Center;
the Ministry of Science and Higher Education of the Russian Federation, Agreement 14.W03.31.0026, 
and the HSE University Basic Research Program, Moscow; 
University of Tabuk research grants
S-1440-0321, S-0256-1438, and S-0280-1439 (Saudi Arabia);
the Slovenian Research Agency Grant Nos. J1-9124 and P1-0135;
Ikerbasque, Basque Foundation for Science, Spain;
the Swiss National Science Foundation; 
the Ministry of Education and the Ministry of Science and Technology of Taiwan;
and the United States Department of Energy and the National Science Foundation.
These acknowledgements are not to be interpreted as an endorsement of any
statement made by any of our institutes, funding agencies, governments, or
their representatives.
We thank the KEKB group for the excellent operation of the
accelerator; the KEK cryogenics group for the efficient
operation of the solenoid; and the KEK computer group and the Pacific Northwest National
Laboratory (PNNL) Environmental Molecular Sciences Laboratory (EMSL)
computing group for strong computing support; and the National
Institute of Informatics, and Science Information NETwork 6 (SINET6) for
valuable network support.

\end{document}

%% file: pub635-orcid_2.tex
\noaffiliation
  \author{S.~Watanuki\,\orcidlink{0000-0002-5241-6628}} 
  \author{G.~de~Marino\,\orcidlink{0000-0002-6509-7793}} 
  \author{K.~Trabelsi\,\orcidlink{0000-0001-6567-3036}} 
  \author{I.~Adachi\,\orcidlink{0000-0003-2287-0173}} 
  \author{H.~Aihara\,\orcidlink{0000-0002-1907-5964}} 
  \author{D.~M.~Asner\,\orcidlink{0000-0002-1586-5790}} 
  \author{H.~Atmacan\,\orcidlink{0000-0003-2435-501X}} 
  \author{V.~Aulchenko\,\orcidlink{0000-0002-5394-4406}} 
  \author{T.~Aushev\,\orcidlink{0000-0002-6347-7055}} 
  \author{R.~Ayad\,\orcidlink{0000-0003-3466-9290}} 
% ...
\collaboration{The Belle Collaboration}